\documentclass[twoside,12pt]{article}
\usepackage{epsfig}

\newcommand{\be}{\begin{equation}}
\newcommand{\ee}{\end{equation}}
\newcommand{\bea}{\begin{eqnarray}}
\newcommand{\eea}{\end{eqnarray}}

\newcommand{\kp}{\mbox{K$^+$}~}
\newcommand{\km}{\mbox{K$^-$}~}

\newcommand{\AGeV}[1][ ]{$A$~GeV{#1}}
\begin{document}

\title{ \vspace{1cm} Strangeness Production at 1 -- 2 $A$ GeV}
\author{H.\ Oeschler \\
\\
Institut f\"ur Kernphysik \\
Darmstadt University of Technology, \\
64289 Darmstadt, Germany \\}

\maketitle
\begin{abstract}
The production of K$^+$ and K$^-$ mesons below and at the NN
threshold is summarized, based on a comparison of data with
transport model calculations. K$^+$ mesons are created in
associate production together with hyperons (e.g.~$\Lambda$) in
multi-step processes involving $\Delta$ resonances. These
processes occur mainly during the high-density phase of the
collision and this makes the K$^+$ an ideal tool to extract the
stiffness of the nuclear equation of state, found to be rather
soft with a compressibility modulus $K$ below 240 MeV. In
contrast, the major part of K$^-$ mesons are produced via
strangeness exchange. Most of the created K$^-$ are absorbed and
the surviving ones are emitted quite late and at low densities.
\end{abstract}
%\eject
%\tableofcontents
\section{Introduction}

Pion production in heavy-ion collisions at a few GeV per nucleon
is the dominant channel for particle emission. Kaon emission,
however, is a rare processes at these energies. In nucleon-nucleon
interactions, the threshold for pion production is 0.29 GeV, for
K$^+$ it is 1.58 GeV and for K$^-$ it is 2.5 GeV. For kaons the
required energy also contains the mass of the lightest associate
partner to be produced in order to conserve strangeness, being a
$\Lambda$ for the K$^+$ and a K$^+$ for the K$^-$. Pions are
predominantly produced in first-chance nucleon-nucleon collisions,
while kaons being created below threshold, require that energy is
accumulated by various means. As a consequence, the K$^+$ mesons
are produced only during the high-density phase, and therefore
these particles are considered as ideal probes for the hot and
dense fireball.

Being produced inside the collision zone, the fate of these
particles is quite different. The pion-nucleon cross section is
large and thus pions are absorbed through the $\Delta$ resonance
and re-emitted by the decay of this resonance. This creation and
disappearance occurs during the entire time evolution of the
collision. The K$^+$-nucleon cross section, on the other hand, is
small, which is due to strangeness and energy conservation. There
are no partners to react with and only elastic scattering and
charge exchange can happen.

The K$^-$ production process is quite different from K$^+$. The
production threshold is much higher. Yet there is an alternative
production channel possible, as suggested by Ko~\cite{Ko_84} and
demonstrated in Refs.~\cite{Oeschler_00,Hartnack_03}. The
strangeness-exchange reaction $ \pi \rm{Y} \rightleftharpoons
\rm{K}^- \rm{N}$ has a large cross section (with Y being $\Lambda$
or $\Sigma$). The inverse channel causes the produced K$^-$ to be
able to be absorbed~\cite{Hartnack_03}. As in the pion case, the
succession of absorption and creation causes the K$^-$ emission to
be in the late stage of the reaction.

In addition, strange hadrons carry information on the properties
of the medium in which they are created and through which they are
propagating due to KN potential interaction~\cite{kaplan}.
%It is also very interesting to consider what
%information on the property of the medium the strange hadrons
%carry. Strange particles interact with the medium not only by
%collisions but also by potential~\cite{kaplan}.
The effective in-medium mass of K$^+$ mesons increases with
density, since the potential is slightly repulsive, while the
K$^-$ potential is strongly attractive and decreases with density.
%It is still an open question what is the best observable to
%extract this information.

Early experiments have been carried out at the Bevalac
accelerator. Systematic, high-statistics measurements of pion and
kaon production became feasible with the advent of the SIS
accelerator at GSI.
%Two experiments, KaoS~\cite{Senger_93}, a dedicated
%experiment for the measurement of K-mesons, and
%FOPI~\cite{Gobbi_93}, a large acceptance, multi-purpose detector,
%produced a solid body of beautiful results. Transport models have
%been developed to study in detail the time evolution of the
%collisions.
In this short review, mainly results from the KaoS
Collaboration~\cite{KaoS} are given, and some examples of recent
IQMD calculations~\cite{iqmd}. A recent review on pion emission
can be found in~\cite{reisdorf_07}.

\section{Particle production yields}

\begin{figure}[htb]
\epsfysize=9.0cm
\begin{center}
\begin{minipage}[t]{7.5 cm}
\epsfig{file=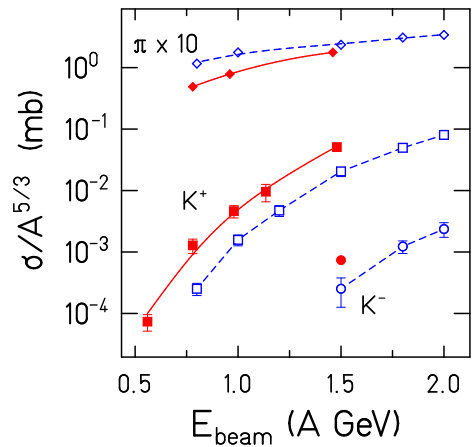,width=7.5cm}
\vspace*{-2.5cm}
\end{minipage}
\vspace*{-.9cm}

\begin{minipage}[t]{16.5 cm}
\caption{Cross sections of pions (diamonds), of K$^+$ (squares),
and of K$^-$ (circles) per mass number $A^{5/3}$ for Au + Au
(filled symbols) and for C + C (open symbols) as a  function of
the beam energy.  Values from   \cite{KaoS}.
\vspace*{-.9cm}
\label{k_sigma_Energy}}
\end{minipage}
\end{center}
\end{figure}

A survey of the production yields of pions and kaons in inclusive
collisions of a light (C+C) and a heavy (Au+Au) system are given
in Fig.~\ref{k_sigma_Energy}  as a function of beam energy where
the ordinate is $\sigma/A^{5/3}$ which represents the multiplicity
per mass number $A$ of the colliding system.  As can be seen from
this figure the pion multiplicity per $A$ is higher in the lighter
system which reflects likely the influence of absorption of pions
and holds only at these low incident energies. In contrast, for
K$^+$ production the inverse observation is made.

\vspace*{-.5cm}
The key mechanism for the K$^+$ production in heavy-ion reactions
close to the threshold is a multi-step process where the energy
necessary for the production is accumulated and stored in
intermediate resonances.  Higher densities increase the number of
these collisions and especially second generation collisions like
$\Delta \rm{N} $ with sufficiently high relative momentum to
create a K$^+$ occur most frequently during the high-density phase
of the reaction. This has nicely been demonstrated by IQMD
calculations~\cite{iqmd} showing that also well above the
corresponding NN threshold the channel $\Delta \rm{N} $ dominates
the K$^+$ production. The yield of \km is about two orders of
magnitude lower due to the much higher threshold.

\begin{figure}[htb]
\begin{center}
\begin{minipage}[t]{14 cm}
\epsfig{file=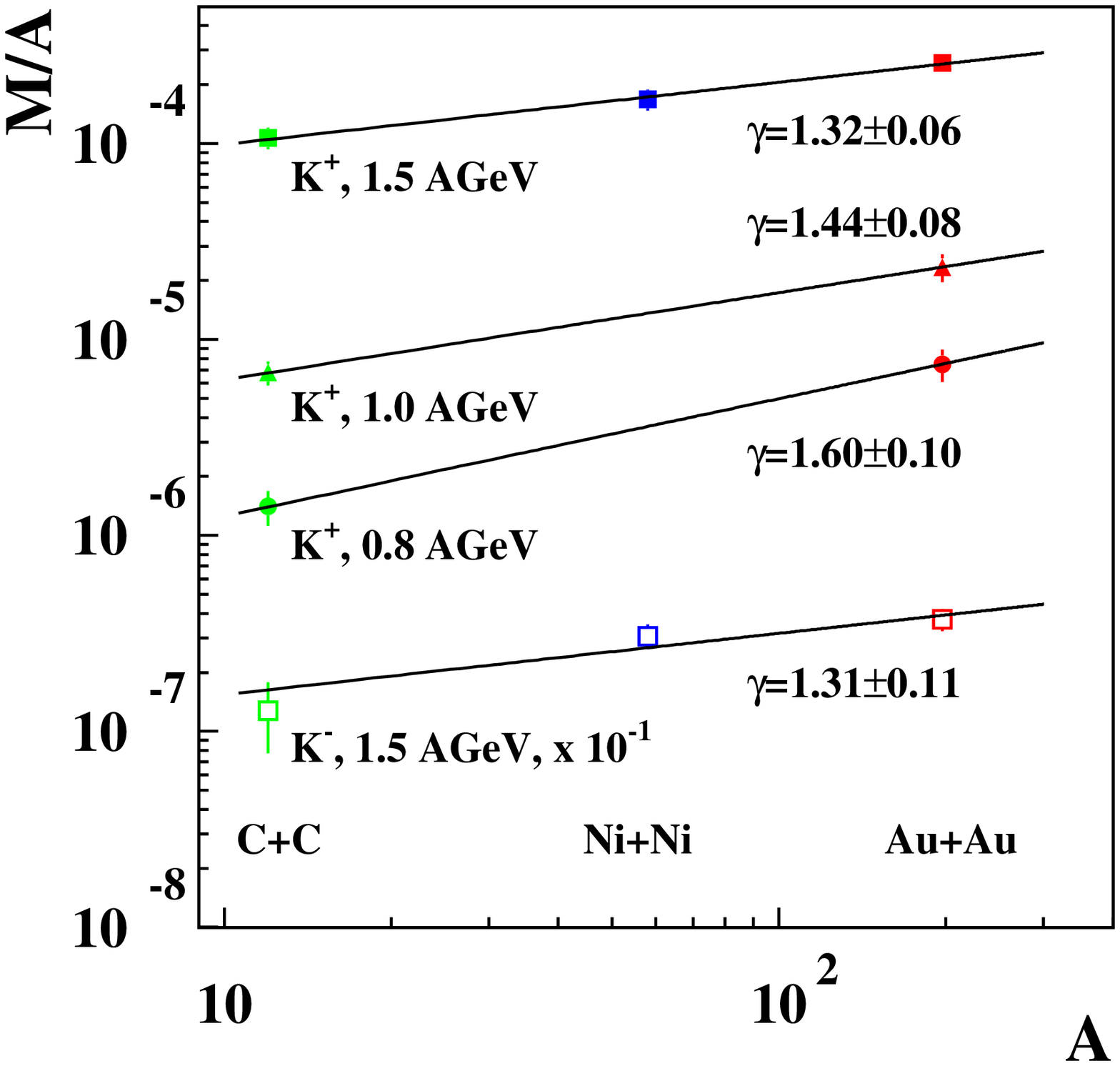,height=7cm}\epsfig{file=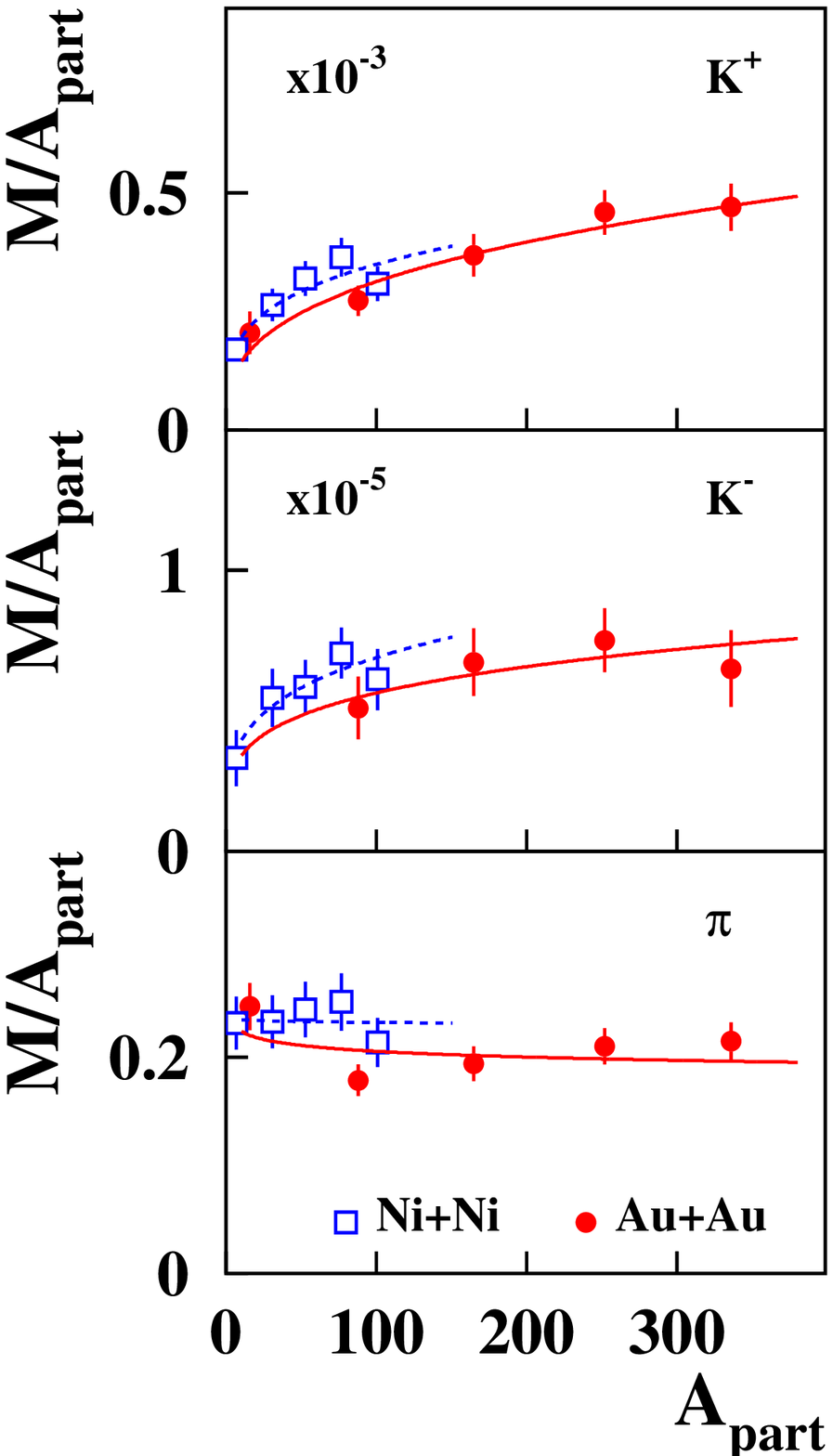,height=7cm}
\end{minipage}
\vspace*{-.8cm}

\begin{minipage}[t]{16.5 cm}
\caption{Left: Multiplicities per mass number $M/A$ as a function
of $A$ of K$^+$ (full symbols) and of K$^-$ (open symbols) for
\mbox{C+C}, \mbox{Ni+Ni}, and \mbox{Au+Au}. The lines represent
the function $M \sim A^{\gamma}$ fitted to the data. Right:
Dependence of the multiplicities of K$^+$ (upper panel) and of
K$^-$ mesons (middle panel) as well as of pions (lower panel) on
$A_{\rm part}$. Full symbols denote \mbox{Au+Au}, open symbols
\mbox{Ni+Ni}, both at $1.5$~\AGeV. The lines are functions $M \sim
A^{\alpha}_{\rm part}$ fitted to the data separately for
\mbox{Au+Au} (solid lines) and \mbox{Ni+Ni} (dashed lines). From
\cite{KaoS}. \label{k_sigma_a}}
\end{minipage}
\end{center}
\end{figure}

Figure~\ref{k_sigma_a}, left, shows the multiplicities of K$^+$
mesons per mass number $A$ from inclusive reactions as a function
of $A$ at several incident energies as well as those of K$^-$
mesons at $1.5$~\AGeV[].
%To interpolate between measured data points in case
%of slight differences in the effective beam energies due to
%different energy losses in the respective targets the fits to the
%excitation functions according to Eq.~(\ref{eq_kolomeitsev}) as
%shown in Figure~\ref{k_sigma_Energy} have been used.
The lines in Fig.~\ref{k_sigma_a} are functions $M \sim
A^{\gamma}$ fitted to the data with the resulting values for
$\gamma$ given in the figure. For K$^+$ production the extracted
values of $\gamma$ decrease with incident energy. This reflects
the decreasing influence of the intermediate energy storage via
$\Delta$ and hence the influence of the density. Considering the
much higher threshold for K$^-$ production, one would expect that
$\gamma$ is much larger at comparable incident energies. However,
at 1.5 \AGeV the values for K$^+$ and for K$^-$ production are
about equal. This confirms that the K$^-$ production is linked to
the K$^+$ production. \km mesons are dominantly created by
strangeness exchange converting a hyperon and a pion into a \km
and a nucleon.

Figure~\ref{k_sigma_a}, right, shows the multiplicity per number
of participating nucleons $M/A_{\rm part}$ for Ni+Ni and Au+Au
collisions at 1.5 \AGeV
 as a function of $A_{\rm part}$ and demonstrates that the
multiplicities per  $A_{\rm part}$ of pions does not change with
$A_{\rm part}$ while both kaon species exhibit the same rise
despite the fact that the thresholds for the production of the two
particles species in binary NN-collisions are very different. This
is observed in \mbox{Au+Au} as well as in \mbox{Ni+Ni} collisions
with comparable multiplicities at the same $A_{\rm part}$. Again,
this observation confirms the strong relation between K$^+$ and
K$^-$ production. \vspace*{-.4cm}
As a consequence of this observation, the \mbox{K$^-$/K$^+$} ratio
as a function of $A_{\rm part}$ is constant as a function of
$A_{\rm part}$ and is even the same for two different systems.
Again, at first glance it is astonishing that in central
collisions with high densities the K$^-$/K$^+$ ratio is the same
as in peripheral ones. Yet, considering the observation from
Fig.~\ref{k_sigma_a}, right, it is expected. Remarkable is that a
statistical model is able to describe these
ratios~\cite{Cleymans_99}.
All these observations indicate that K$^-$ are essentially created
by the strangeness-exchange mechanism and it has been shown by
applying the law-of-mass action that this channel might reach
chemical equilibrium~\cite{Cl04}.

\section{Spectra}

\begin{figure}[htb]
\begin{center}
\begin{minipage}[t]{13.2 cm}
\epsfig{file=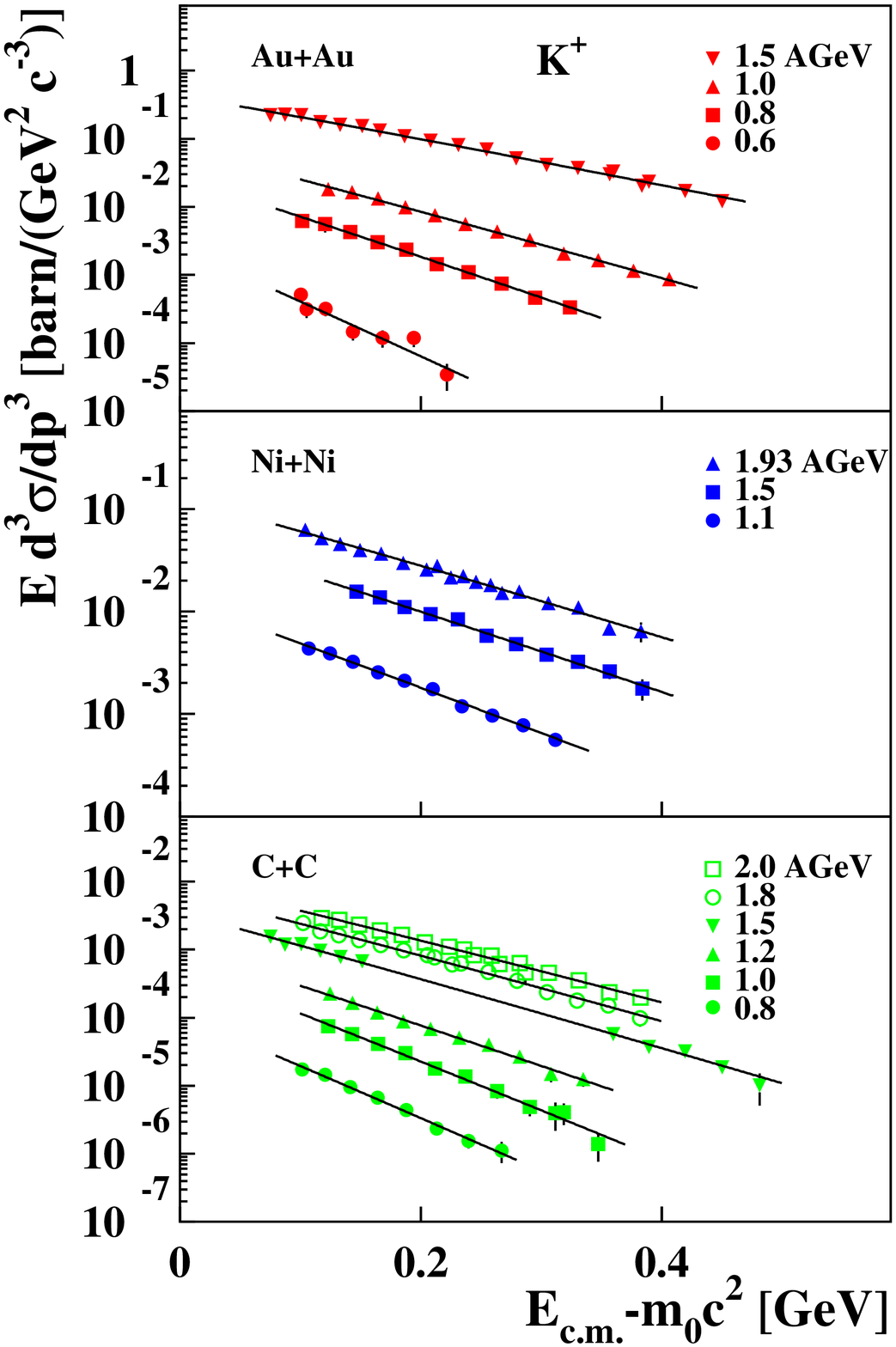,width=6cm}
\epsfig{file=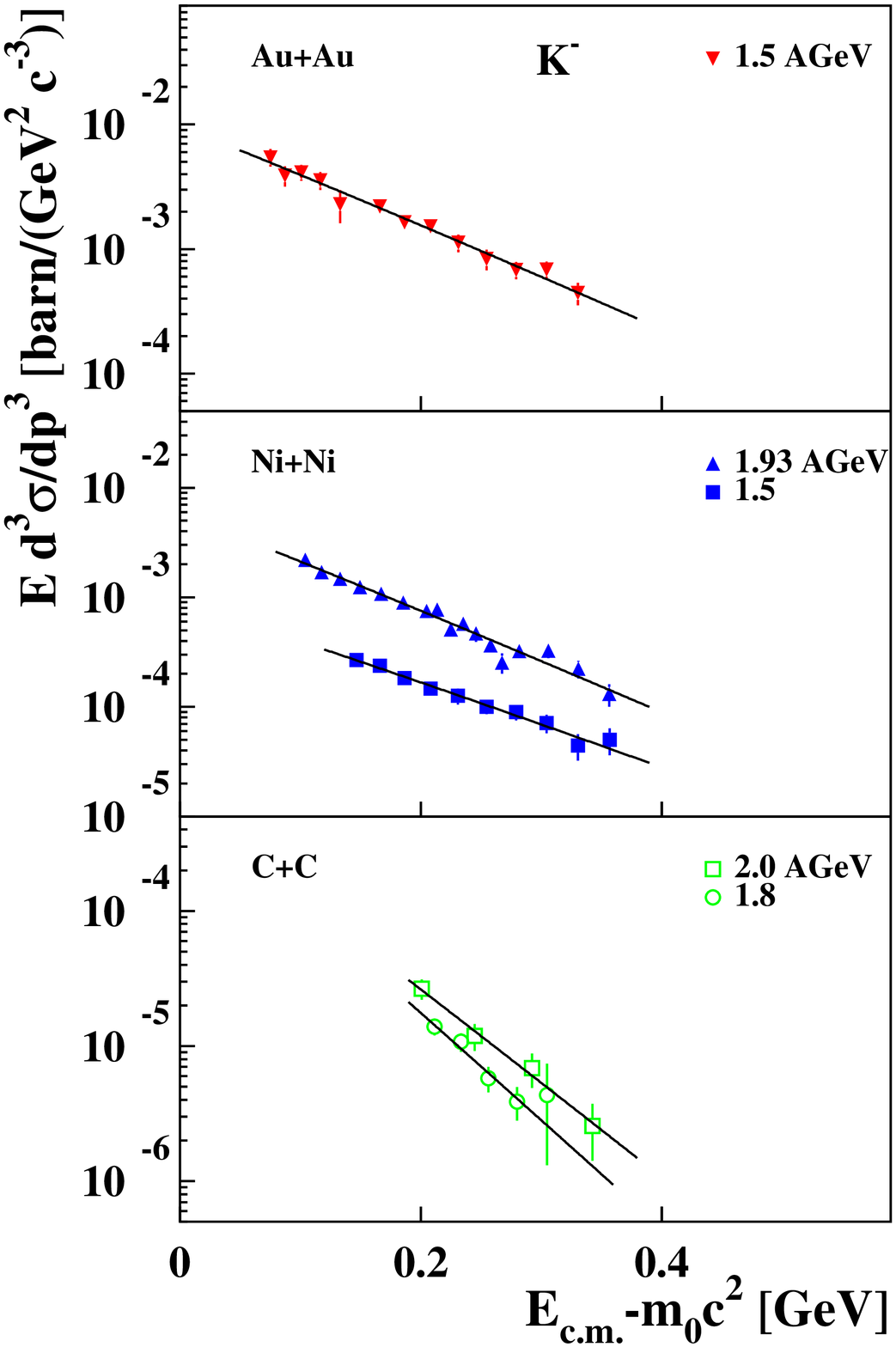,width=6cm}
\end{minipage}
\vspace*{-.5cm}

\begin{minipage}[t]{16.5 cm}
\caption{Inclusive invariant cross sections as a function of the
kinetic energy $E_{\rm c.m.} - m_{0}c^{2}$ for K$^+$ (left) and
for K$^-$ (right) for three collision systems and various beam
energies at mid-rapidity ($\theta_{\rm c.m.} = 90^\circ \pm
10^\circ$). From \cite{KaoS}. \vspace*{-.7cm} \label{midrap}}
\end{minipage}
%\vspace*{-.5cm}
\end{center}
\end{figure}

Both K$^+$ and K$^-$ mesons are produced in a complicated sequence
of interactions. Figure~\ref{midrap} shows K$^+$ and K$^-$ spectra
at midrapidity as a function of the kinetic energy $E_{\rm c.m.} -
m_{0}c^{2}$ for three different systems  and various beam
energies. The spectra have a Boltzmann shape to a very good
approximation. Two general trends can be seen from this figure:
(i) The slopes of the K$^+$ are always higher than those of K$^-$.
(ii) Heavy systems exhibit higher slopes than lighter ones.

\begin{figure}[tb]
\begin{center}
\begin{minipage}[t]{10.5 cm}
\epsfig{file=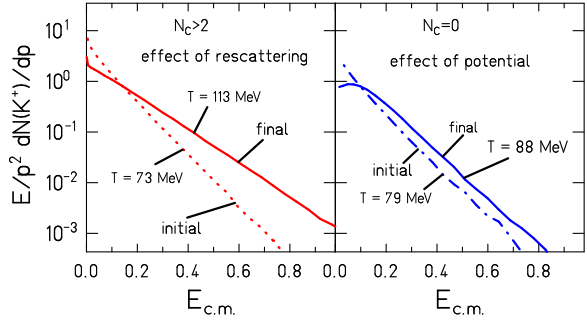,height=5cm}
\end{minipage}
\vspace*{-.7cm}

\begin{minipage}[t]{16.5 cm}
\caption{Influence of rescattering of the K$^+$ and of the
repulsive KN potential demonstrated for central Au+Au collisions
at 1.5 $A$ GeV. Left: Influence of the rescattering of K$^+$
mesons by selecting kaons which have scattered twice or more ($N_C
>$ 2), showing their initial and final distribution. Right:
Influence of the KN potential on the spectral shape demonstrated
by selecting kaons that never scattered ($N_C$ = 0) and comparing
the initial and final spectra~\cite{iqmd}. Without potential the
initial spectrum is also the final one.\vspace*{-.7cm}
\label{spectra-effects}}
\end{minipage}
\end{center}
\end{figure}

It is interesting to study the different components and effects
contributing to the K$^+$ slope in an IQMD study~\cite{iqmd}.
Initially, the energy available for K$^+$ production is low and
the slope at creations is therefore quite steep as can be seen
from Fig.~\ref{spectra-effects}. The final slopes are influenced
by re-scattering of the K$^+$ in the medium and by the repulsion
from K$^+$N potential as demonstrated in this figure.
Re-scattering dramatically increases the slope. This effect
increases as the beam energy increases and as the mass of the
system increases as there is more scattering in heavier systems.
The expected repulsive K$^+$ potential increases the production
threshold and might modify the spectra only at low momenta which
are difficult to detect due to the kaon decay. For experimental
reasons kaons at low momenta can be studied best with neutral
K$^0_s$ mesons~\cite{Merschmeyer_07,HADES_AS}.
%
%\vspace*{-.8cm}
One might be astonished that scattering contributes so strongly as
the mean free path of K$^+$ is about 5 fm. Yet, K$^+$ are produced
at about twice normal nuclear matter density and mostly inside the
collision zone. Hence, the mean free path decreases drastically
and K$^+$ have to travel a distance before leaving the system as
shown in~\cite{iqmd}.

\begin{figure}[htb]
\begin{center}
\begin{minipage}[t]{16 cm}
\epsfig{file=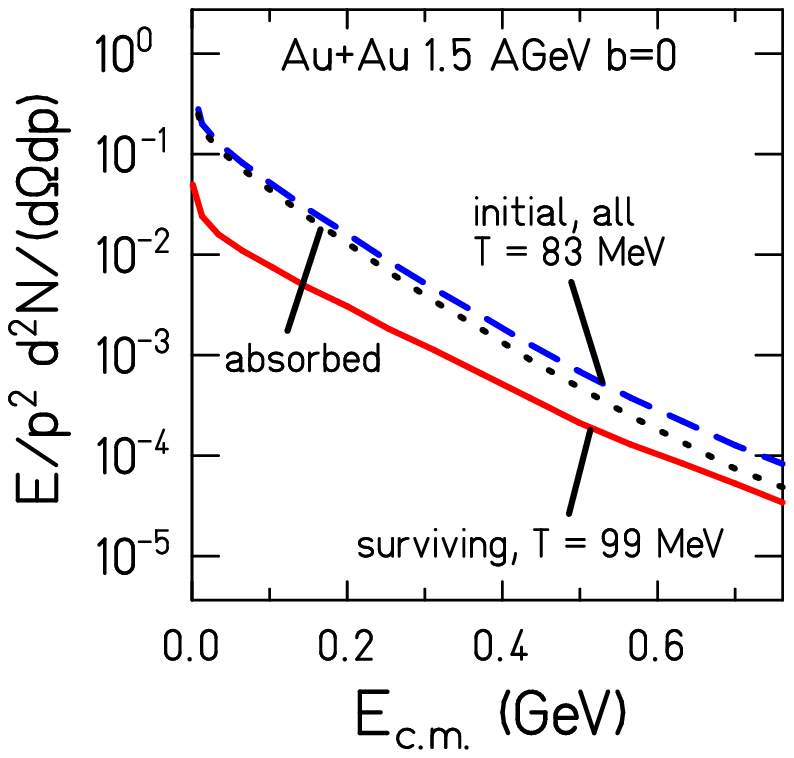,height=5cm}
\epsfig{file=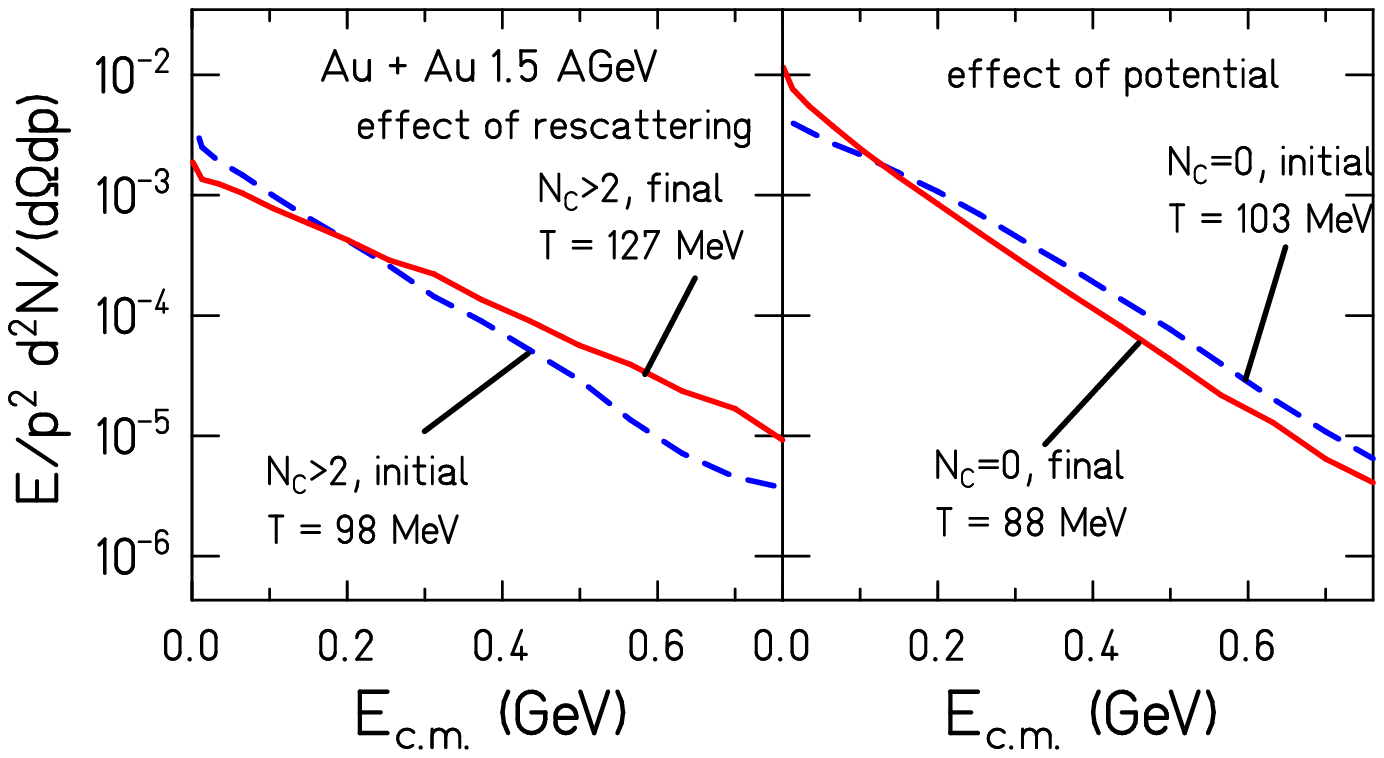,height=5cm}
\end{minipage}
\vspace*{-.7cm}

\begin{minipage}[t]{16.5 cm}
\caption{Left: Influence of absorption of \km mesons demonstrated
for \mbox{Au+Au} at $1.5$~\AGeV based on IQMD calculations. Right:
Effect of rescattering of K$^-$ and of the attractive KN
potential~\cite{iqmd}. \vspace*{-.7cm} \label{spec_abs_nc_pot}}
\end{minipage}
\end{center}
\end{figure}

 The slope of the K$^-$ is influenced by scattering the same way as the
 K$^+$, yet scattering occurs less frequently since absorption dominates when K$^-$ interact with nuclei. According to
 IQMD calculations only about 20\% of the K$^-$  produced in Au+Au finally leave the system as shown in Fig.~\ref{spec_abs_nc_pot}.
 Furthermore, due to the momentum dependence of the absorption, the low momenta of the K$^-$ undergo
 stronger absorption which leads to an increase in slope or apparent
 temperature. The influence of the KN potential is only visible at
 low momenta which are very difficult to measure.
%
%\vspace*{-.8cm}

\begin{figure}[htb]
\begin{center}
\begin{minipage}[t]{8.5 cm}
\epsfig{file=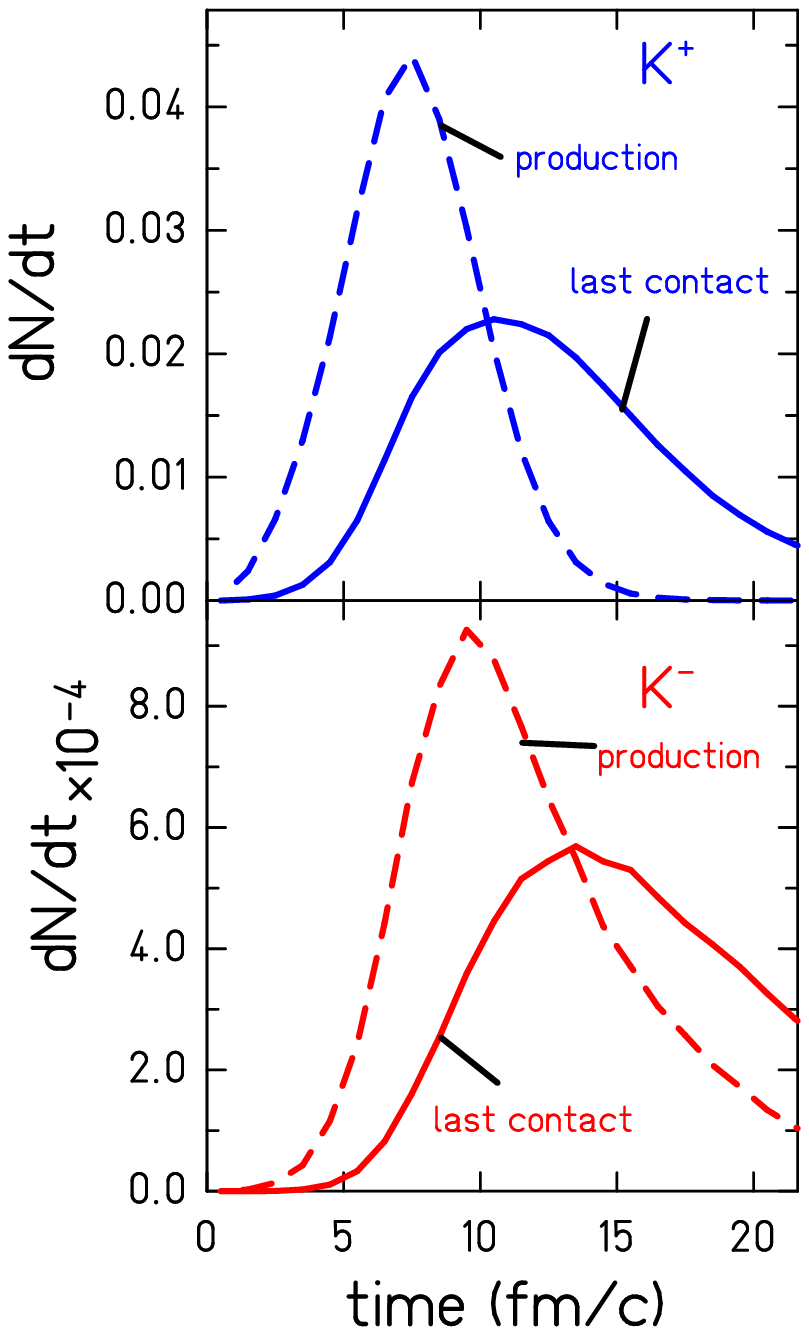,width=4cm}
\epsfig{file=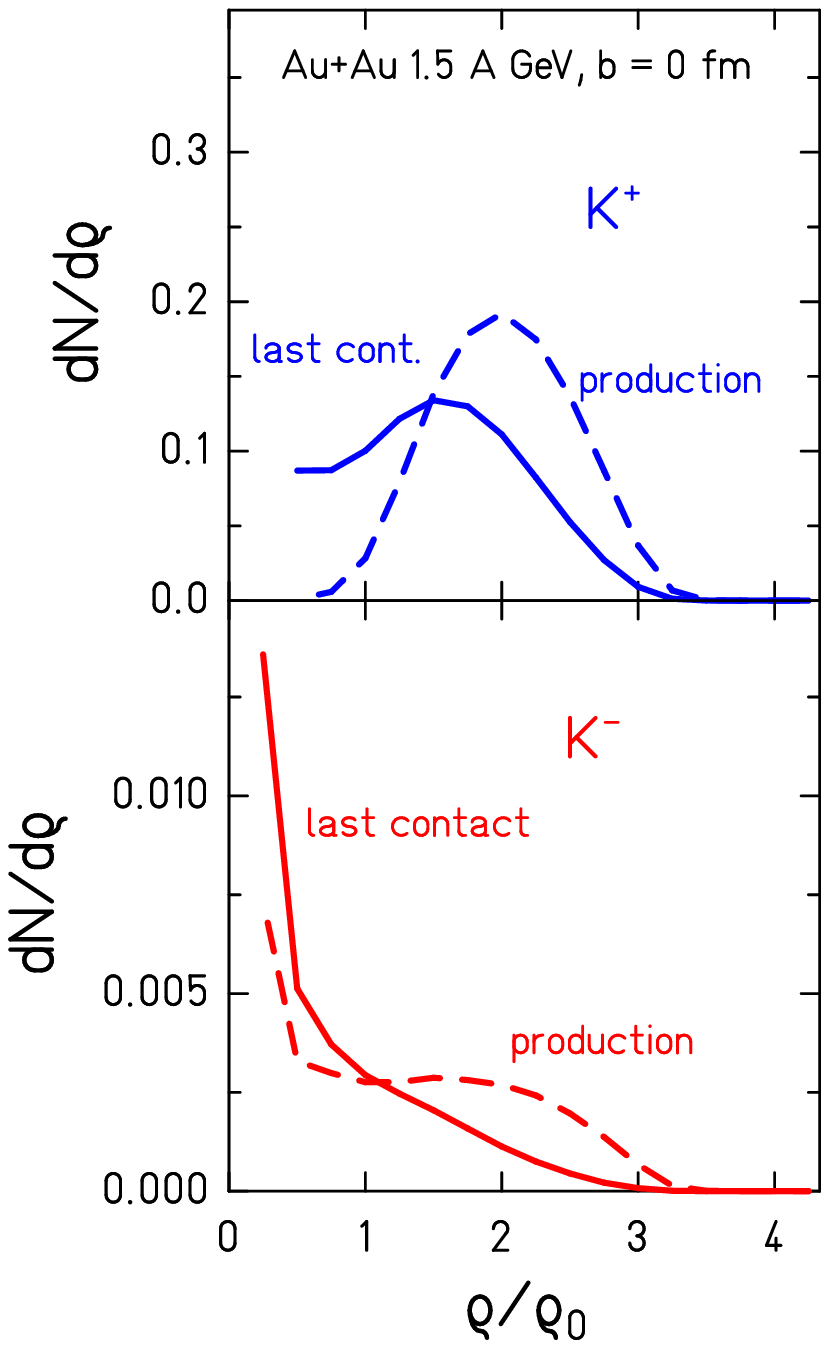,width=4cm}
\end{minipage}
\vspace*{-.7cm}

\begin{minipage}[t]{16.5 cm}
\caption{Left: Time profiles for production (dashed lines) and
 last contact (solid lines) of K$^+$ (top) and K$^-$ (bottom) mesons.
 Right: Density distribution of the medium  at the point of production (dashed lines)
 and at the point of last contact (solid lines) for K$^+$ (top) and K$^-$ (bottom) mesons.
 The simulations are for central $Au + Au$ collisions at 1.5 $A$ $GeV.$~\cite{iqmd}.
\vspace*{-.7cm}
 \label{density-time}}
\end{minipage}
\end{center}
\end{figure}

In order to put these various observables in relation to the
expected in-medium properties of kaons, it is useful to study the
emission time of the two kaon species and the corresponding
density profiles. Figure~\ref{density-time} shows on the left side
the time distributions of kaons at the point of production and at
the time of last contact. As discussed before, K$^-$ are produced
later than K$^+$ and they also leave the system later. The right
hand side of Figure~\ref{density-time} shows the corresponding
density distributions. The bulk of the K$^+$ is produced when the
density is twice nuclear matter density and
%they are emitted predominantly at
%a density of $\rho$ = $1.5$ $\rho_0$.
their yield cannot be changed between production and emission due
to strangeness conservation. This fact is the key property as to
why the \kp gives access to the high-density phase and thus allows
one to extract the nuclear equation of state. In contrast, the
K$^-$ are predominantly produced and emitted later and from a
region of density below nuclear matter density. From this model
study, it is obvious that heavy-ion reactions in the 1 to 2 \AGeV
regime are well suited to also study K$^+$ potential effects, but
they are not likely to yield results on K$^-$ potential effects
since those effects are expected to be small at densities below
nuclear matter density.
\vspace*{-.5cm}

\section{The Nuclear Equation of State}

One of the challenging questions of nuclear physics is the
determination of the nuclear equation of state (EoS). One of the
very successful strategies refers to the study of monopole
vibrations which yields a value for the compression modulus $K$ of
235$\pm$ 14 MeV~\cite{Youngblood} with $K = -V\frac{{\rm d}p}{{\rm
d}V}= 9 \rho^2 \frac{{\rm d}^2E/A(\rho,T)}{({\rm d}\rho)^2}
|_{\rho=\rho_0}$ which measures the curvature of $E/A(\rho,T)$ at
the equilibrium point. This information is however, limited to
tiny density variations around $\rho_0$. Heavy-ion collisions are
ideally suited to reach high densities and they might allow us to
extract further information on the EoS. In semi-central heavy-ion
collisions, an in-plane flow is created due to the transverse
pressure on the baryons outside of the interaction region with
this flow being proportional to the transverse pressure. These
studies have not yet lead to clear results~\cite{ant} and earlier
conclusions~\cite{paw2} have to be taken with care. The most
promising method seems to be the strangeness production and its
sensitivity to the energy available during the hot and compressed
phase as first pointed out in~\cite{aik}.

\begin{figure}[htb]
\begin{center}
\begin{minipage}[t]{16 cm}
\epsfig{file=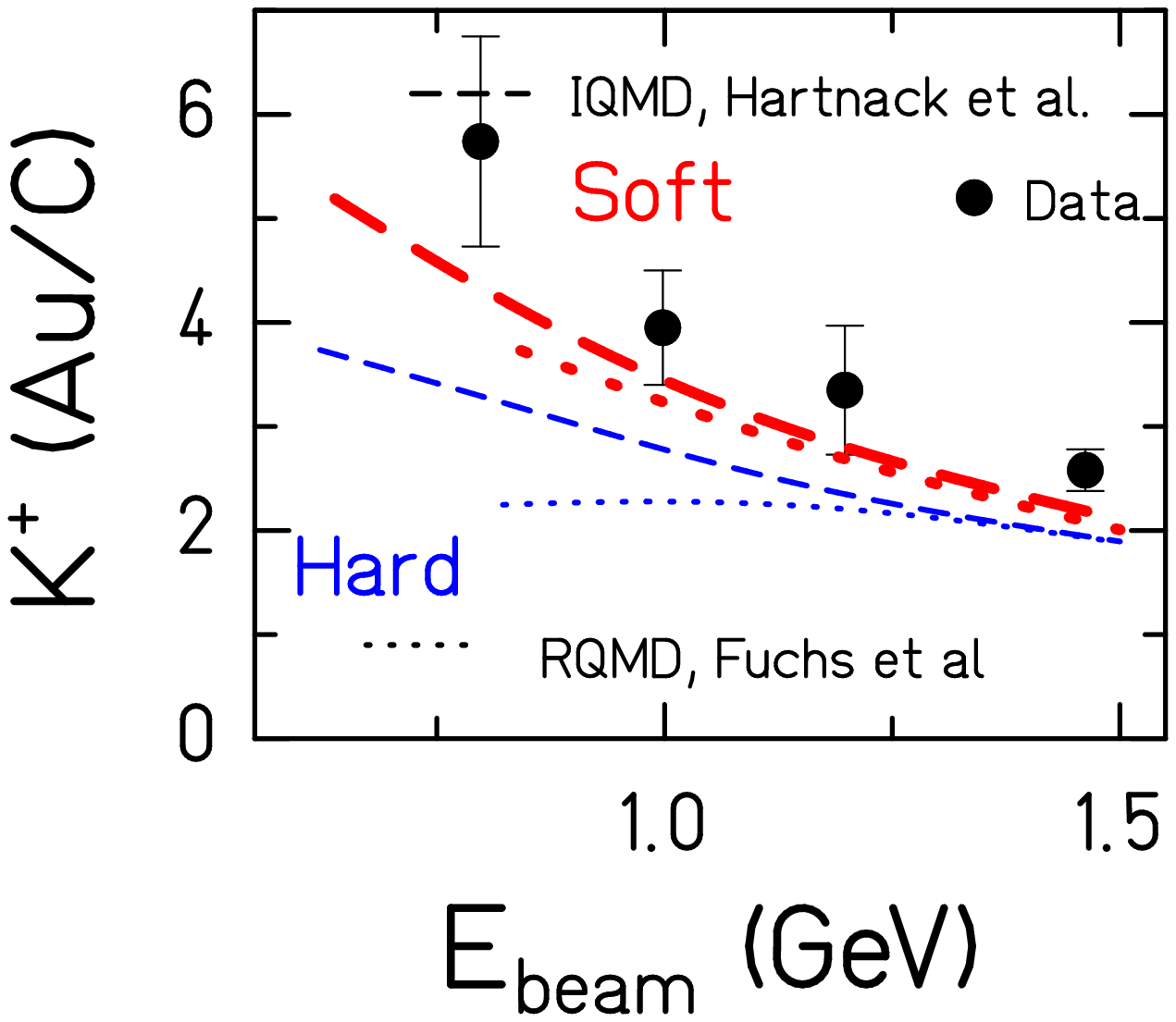,height=5cm}
\epsfig{file=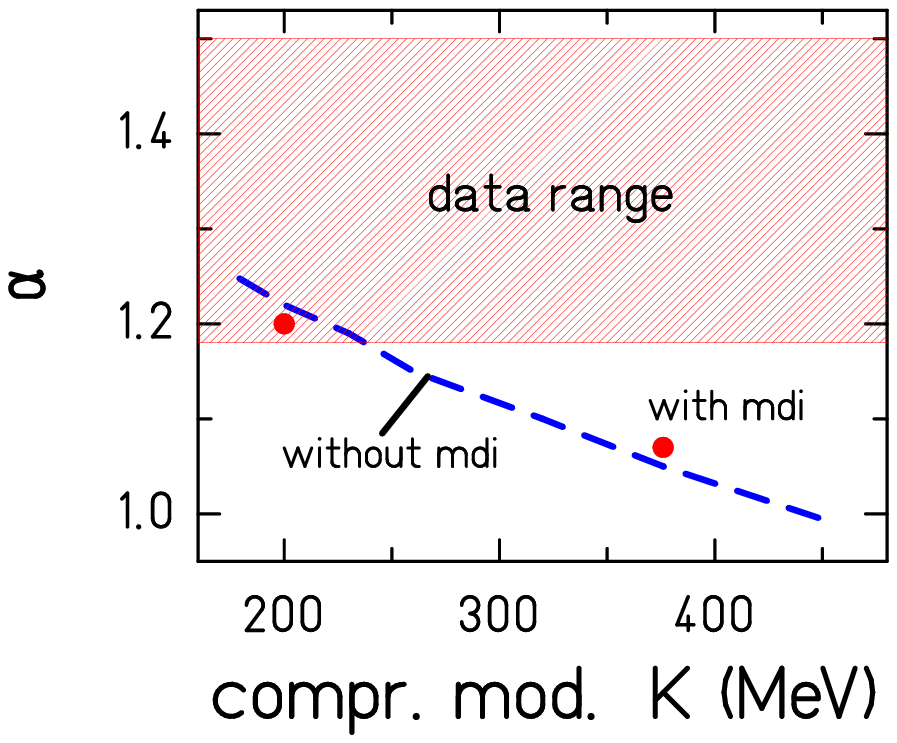,height=5cm}
\end{minipage}
\vspace*{-.7cm}

\begin{minipage}[t]{16.5 cm}
\caption{Left: Comparison of the experimental $K^+$ excitation
functions \cite{sturm} of the
  double ratio $(M_{\mbox{K$^+$}}/A)_{Au+Au} / (M_{\mbox{K$^+$}}/A)_{C+C}$ (the K$^+$ multiplicities per mass number $A$) obtained
  in \mbox{Au+Au} divided by the one in \mbox{C+C} with
  RQMD~\cite{fuchs} (dotted) and with
  IQMD calculations~\cite{Hart_eos} (dashed). Comparing the results of a
  soft (bold red) with a hard (thin blue) EoS. Right: The exponent $\alpha$ is shown as a
function of the compression modulus $K$ determined from IQMD
calculations (with and without momentum-dependent interactions,
mdi) and compared to the values extracted from Au+Au collisions at
1.5 $A$ GeV (Fig.~2 right). \label{EOS}} \vspace*{-.7cm}
\end{minipage}
\end{center}
\end{figure}

However, a comparison of the K$^+$ excitation function with model
calculations suffers from uncertainties in the input quantities
and also on the unknown influence of the K$^+$ N potential. It has
been realized that many of the theoretical and experimental
uncertainties disappear if ratios of K$^+$ multiplicities are
used~\cite{sturm}. The double ratio $
 (M_{\mbox{K$^+$}}/A)_{Au+Au} / (M_{\mbox{K$^+$}}/A)_{C+C} $
turns out to be directly sensitive to the stiffness of EoS. In C+C
collisions the densities hardly exceed $\rho_0$ providing a good
normalization to cancel out systematic uncertainties both in the
experimental data and in unknown input quantities of the transport
model calculations. This ratio is plotted in Fig.~\ref{EOS} as a
function of the beam energy for a soft (bold red lines) and a hard
(thin blue) EoS. The dotted lines refer to RQMD calculations by
Fuchs \cite{fuchs} and both calculations agree quite well. This
figure elucidates again the sensitivity of the EoS being higher at
the lowest beam energies, and demonstrates that only a soft EoS is
compatible with the experimental data of the KaoS
collaboration~\cite{sturm,fuchs}.

%It is, however,  not sufficient that one setup of the simulation
%programs shows agreement with data. Figure~\ref{EOS} shows both
%the results from RQMD and from IQMD and they agree quite well.
Furthermore, in order to have a robust observable one has to
demonstrate that the uncertainties of the input variables do not
render the conclusion useless. Therefore, detailed calculations
with different $N+\Delta \to \mbox{K$^+$} NN$ cross section, with
and without KN potential and for different lifetimes of the Delta
resonance have been performed and none of these uncertainties is
able to weaken the previous conclusion that only a soft EoS is
compatible with the observed excitation function of the K$^+$
yield~\cite{Hart_eos}.

Can this conclusion on the EoS be confirmed independently? Instead
of varying the size of the system one can also vary the
centrality. Because a higher centrality of the collisions yields a
higher compression we expect that the K$^+$ yield per participant
as a function of the centrality depends as well on the EoS.
The measured dependence of  $M/A_{\rm part}$ with $A_{\rm part}$
exhibits a rise $\propto A_{{\rm part}}^\alpha$ shown in
Fig.~\ref{k_sigma_a}, right. The extracted value of $\alpha =
1.34\pm0.16$ for Au+Au collisions at 1.5 \AGeV~\cite{KaoS}  is
shown, as the band in Fig.~\ref{EOS}, right. The dashed line shows
IQMD calculations and also for this observable the data of the
KaoS collaboration are only compatible with values of K around 200
MeV. Thus two independent observables point towards a rather low
compression modulus.

\section{Conclusion}

The two kaon species exhibit quiet distinct differences when
produced in heavy-ion reactions below or at the threshold in NN
collisions.

\kp mesons are created in associate production together with
hyperons. This happens mainly in multi-step processes via the
intermediate $\Delta$ resonance. According to transport model
calculations, they are produced early -- around 7 fm/c -- and at
densities around twice normal nuclear matter. Before leaving the
collision zone, they scatter quite often which causes their
spectral slopes to be increased, but not changing their yield.

\km mesons, in contrast, are mainly created by strangeness
exchange $ \pi \rm{Y} \rightleftharpoons \rm{K}^- \rm{N}$ but they
are as well easily reabsorbed by the inverse reaction and those
finally emitted test the late phase of the collision.

Hence, the production yields of \kp mesons have been used to
extract the stiffness of the nuclear equation of state. Hints for
a repulsive \kp N potential have been found, but this remains an
open issue. Experimental hints for an attractive \km N potential
exist, but are still strongly debated.

\end{document}